\documentclass[prc,aps,showpacs,showkeys,groupedaddress,floatfix,superscriptaddress]{revtex4-1}
\usepackage{epsfig}
\usepackage{bm}
\usepackage{graphics}
\usepackage{graphicx}
\usepackage{epsfig}
\usepackage{amssymb}
\usepackage{amsmath}
\usepackage{color}
\usepackage{natbib}

\begin{document}

\title{Effect of the modified Heisenberg algebra on the Black hole
thermodynamics}

\author{Bilel Hamil}
\email{hamilbilel@gmail.com}
\affiliation{D\'{e}partement de TC de SNV, Universit\'{e} Hassiba Benbouali, Chlef, Algeria.}

\author{Bekir Can L\"{u}tf\"{u}o\u{g}lu}
\email{Corresponding author : bclutfuoglu@akdeniz.edu.tr}

\affiliation{Department of Physics, Akdeniz University, Campus 07058 Antalya, Turkey}
\affiliation{Department of Physics,  University of Hradec Kr\'{a}lov\'{e}, Rokitansk\'{e}ho 62, 500\,03 Hradec Kr\'{a}lov\'{e}, Czechia}

\date{\today}

\begin{abstract}		
In this letter, we introduce two new forms of uncertainty relations by extending the usual Heisenberg algebra with higher terms. Within these scenarios, at first, we study on the Unruh temperature.  We find that the Unruh temperature  increases in one of the new forms while it decreases in the other form. Next, we explore the thermodynamic properties of a Schwarzschild black hole by relying on these two new forms of the Heisenberg uncertainty relation. We present the modifications in the mass-temperature, specific heat, and entropy functions of the black hole according to the extension parameters.
\end{abstract}

\maketitle

\section{Introduction} It is a well known fact that  to make an "infinitesimal" length measurement is impossible \cite{Garay, Hossenfelder}. The smallest measurable fundamental length, hereafter ML, is known to be in the Planck length scale, that is $L_P=\sqrt{\hbar G / c^3} \approx 1.6 \times 10^{-35} \mathrm{m}$ \cite{MigUkr}. In quantum mechanics according to the Heisenberg uncertainty principle (HUP) another fundamental lower limit exists in the measurement of certain physical quantity pairs simultaneously, i.e. $\Delta x \Delta p \geq \hbar/2$. However, for a  very well localized particle, the uncertainty in its momentum could be very high, in so doing, the uncertainty in its position becomes infinitesimally small. Therefore, it is said that the HUP, thus, ordinary quantum mechanics does not predict a ML value. In order to include a smallest measurable fundamental length concept to the quantum mechanics, a modification to the usual Heisenberg algebra is being suggested \cite{Kempf1995}. However, it is shown that there is not a unique deformation formalism of the algebra \cite{Kempf2001, Kempfetal2001, Changetal20022}.  In the last decades, many researches that addressed various deformations in the Heisenberg algebra  \cite{Kempf1995, Kempf2001, Kempfetal2001, Changetal20022, Nouicer2007, Alietal2009, Dasetal2010, Alietal2011, Pedram1, Pedram2, HC1, HC2, SC, Po, Ovgun2016,  Ovgunetal2016, Ovgunetal2017, Rami, Dabrowski, Hassan}. Some of them, i.e. Kempf-Mangano-Mann (KMM) \cite{Kempf1995}, Nouicer (Nr) \cite{Nouicer2007}  foresee solely a ML, while some of the others, i.e.  Pedram (Pm) \cite{Pedram1, Pedram2}, Hassanabadi-Chung (HC) \cite{HC1, HC2}, Shababi-Chung (SC) \cite{SC}, Petruzziello (Po) \cite{Po} predict a maximal observable momentum value as well as the ML uncertainty.

In fact, the basic idea of the presence of a ML was given much earlier by Snyder. He had introduced a noncommutative geometry with the deformation of commutation relations in order to get rid of the divergences in quantum field theory \cite{Snyder1, Snyder2}.
\begin{eqnarray}
&&\left[ X_{\mu },P_{\nu }\right] =i\hbar \left( \eta _{\mu \nu }+\theta P_{\mu
}P_{\nu }\right) ;\text{ \ } \nonumber \\
&&\left[ X_{\mu },X_{\nu }\right] =i\hbar \theta
J_{\mu \nu };\text{ \ }\left[ P_{\mu },P_{\nu }\right] =0,
\end{eqnarray}%
where, $\eta_{\mu \nu }$ denotes the metric tensor while its signature is given by $\left[ \eta_{\mu \nu }\right] =\mathrm{diag} \left(
\begin{array}{cccc}
-1 & 1 & 1 & 1%
\end{array}%
\right) $. Here, $\theta $ is a deformation parameter that is proportional to the square of the  Planck length.  $J_{\mu \nu }=x_\mu p_\nu- x_\nu p_\mu$ are the generators of Lorentz transformations with the spacetime indices $\mu $ and $\nu $.  The model differs fundamentally according to the sign of the deformation parameter.
For instance, when $\theta>0$, the model possesses a discrete spatial and a continuous time spectrum. However, for $\theta<0$,  the model starts to own a continuous spatial and a discrete time spectrum. Thus, to distinguish them from each other, the latter form is  called as anti-Snyder model.

Snyder's model was very brilliant especially in the context of preserving the full Poincaré invariance, however, the model did not gather enough attention in those days.
This indifference to the Snyder model can only be understood from the perspective of those days: "Evolving renormalization techniques are eliminating divergences. Therefore, is there a real need to define a minimum length?". However, the latest developments, especially the newborn string theory and some novel approaches to the quantum gravity theory  showed that the definition of a minimal length is an inevitable necessity \cite{Konoshi, Gross, Magg, Witt}.

In the last decade, we see the reuse of the Snyder model in the context of modified Heisenberg algebra in cosmological and field theoretical researches. For example, Battisti and very recently Mignemi  proposed scaled non-commutative coordinates with different choices of isotropic  momentum-dependent functions \cite{battisti1,battisti2, mignemi}. We intend to use these modified Heisenberg algebra of the Snyder model to investigate a ML and then study some applications according to them. However, during the preparation of this letter, Petruzziello
presented his work by using one of the modified algebra \cite{Po}. In his paper, he used the anti-Snyder model deformation, which has a different fundamental base, and obtained a maximal observable momentum. We note that in the Snyder's model we do not end up with a such value, therefore our findings are novel in the field.

Therefore, in this letter with this given motivation we consider two non-equivalent modified Heisenberg algebras of the one dimensional Snyder model. We generalize the Heisenberg algebra by changing the position operator in the first and the momentum operator in the second cases. For both cases, we obtain a ML value and compare it with the ones that exist in the literature. Then, we give the heuristic derivation of the Unruh temperature according to these modified algebras. After all, we examine the thermodynamic functions of a Schwarzschild black hole in particular in the context of the mass-temperature, specific heat and entropy functions and conclude the paper.

\section{Modified generalized uncertainty principle (MGUP)}
First, we consider the deformed commutation relation between the operators of the position, $X$, and momentum, $P$, in the modified Snyder model that is presented in \cite{mignemi,battisti1,battisti2}.
\begin{equation}
\left[ X,P\right] =i\hbar \sqrt{1+\alpha P^{2}}.  \label{a}
\end{equation}
Here, $\alpha =\alpha _{0}/M_{p}c^{2}=\alpha _{0}L _{p}/\hbar $,  where $M_{p}$ and $L _{p}$ are the Planck mass and length, respectively.  Note that in the limit of  $\alpha
\rightarrow 0$, it reduces to the ordinary Heisenberg algebra.  In order to assure the above modified commutation relation, the position and momentum operators  can be considered  in the momentum space representation as follows:
\begin{equation}
P=p; \quad X=i\hbar \sqrt{1+\alpha p^{2}}\frac{\partial}{\partial p}.
\end{equation}
Please note that, the position and momentum operators  are Hermitian operators according to the following scalar product
\begin{equation}
\left \langle \varphi \right. \left \vert \psi \right \rangle =\int \frac{dp%
}{\sqrt{1+\alpha p^{2}}}\varphi ^{\ast }\left( p\right) \psi \left( p\right).
\end{equation}%
Next, we check whether the modified algebra results with a ML or not. By expanding Eq. (\ref{a}) with respect to the deformation parameter, we get
\begin{equation}
\left[ X,P\right] =i\hbar \left( 1+\frac{\alpha }{2}P^{2}-\frac{\alpha ^{2}}{%
8}P^{4}+\frac{\alpha ^{3}}{16}P^{6}-\frac{5\alpha ^{4}}{128}P^{8}+...\right).
\end{equation}%
We obtain the modified uncertainty relation, namely MGUP, as
\begin{eqnarray}
\left( \Delta X\right) \left( \Delta P\right) &\geqslant &\frac{\hbar }{2}%
\left \langle \sqrt{1+\alpha p^{2}}\right \rangle ,  \notag \\
&\geqslant &\frac{\hbar }{2}\left \langle 1+\frac{\alpha }{2}P^{2}-\frac{%
\alpha ^{2}}{8}P^{4}+\frac{\alpha ^{3}}{16}P^{6}-...\right \rangle ,  \notag
\\
&\geqslant &\frac{\hbar }{2}\left( 1+\frac{\alpha }{2}\left \langle
P^{2}\right \rangle -\frac{\alpha ^{2}}{8}\left \langle P^{4}\right \rangle +...\right).
\label{b}
\end{eqnarray}
By employing the property $\left \langle P^{2n}\right \rangle \geqslant
\left
\langle P^{2}\right \rangle ^{n}$ \cite{HC1,HC2, SC, Po} , we find that Eq. (\ref{b}) becomes%
\begin{eqnarray}
\left( \Delta X\right) \left( \Delta P\right) &\geqslant& \frac{\hbar }{2}%
\bigg( 1+\frac{\alpha }{2}\left[ \left( \Delta P\right) ^{2}+\left \langle
P\right \rangle ^{2}\right] \nonumber \\
&-&\frac{\alpha ^{2}}{8}\left[ \left( \Delta
P\right) ^{2}+\left \langle P\right \rangle ^{2}\right] ^{2}\nonumber \\
&+&\frac{\alpha
^{3}}{16}\left[ \left( \Delta P\right) ^{2}+\left \langle P\right \rangle
^{2}\right] ^{3}-... \bigg).
\end{eqnarray}
At the first order $\alpha $, we find%
\begin{equation}
\left( \Delta X\right) \left( \Delta P\right) \geqslant \frac{\hbar }{2}%
\sqrt{1+\alpha \left[ \left( \Delta P\right) ^{2}+\left \langle P\right
\rangle ^{2}\right] }.  \label{b1}
\end{equation}%
Therefore, we write the absolute smallest uncertainty in position in the form of
\begin{equation}
\left( \Delta X\right) _{MGUP}\simeq \hbar \sqrt{\frac{\alpha }{2}}. \label{b1a}
\end{equation}%
After a  comparison of this finding with the KMM model's result, we conclude that we obtain a smaller ML value, thus, $\left( \Delta X\right) _{MGUP}<\left( \Delta X\right)
_{KMM}$ \cite{kempf,kempf1,kempf2}.

\section{Modified extended uncertainty principle (MEUP)}
As the second deformed algebra, we consider the following modified commutation relation between the
position and momentum operators \cite{Nasrin} :%
\begin{equation}
\left[ X,P\right] =i\hbar \sqrt{1+\beta X^{2}},  \label{c}
\end{equation}
where $\beta =\frac{1}{l_{H}^{2}}$ and $l_{H}^{2}$ is (anti)-de Sitter radius. In the position representation, the following realization of the operator satisfies
the above modified commutation relation.
\begin{equation}
X=x; \quad P=\frac{\hbar }{i}\sqrt{1+\beta x^{2}}\frac{\partial }{\partial x}.
\end{equation}%
Note that in this case of MEUP,  position representation operators act on the square integrable
functions%
\begin{equation}
\psi \in \mathcal{L}\left( \mathbf{R,}\frac{dx}{\sqrt{1+\beta x^{2}}}%
\right) ,
\end{equation}
Moreover, these operators are symmetric operators and they obey  the following inner product rule
\begin{equation}
\left \langle \psi \right. \left \vert \chi \right \rangle =\int \frac{dx}{%
\sqrt{1+\beta x^{2}}}\varphi ^{\ast }\left( x\right) \chi \left( x\right).
\end{equation}%
Furthermore, the scalar product between position eigenstates appears modified as $\left \langle x\right. \left \vert x^{\prime }\right \rangle =\sqrt{1+\beta
x^{2}}\delta \left( x-x^{\prime }\right) .$ Next, we show that Eq. (\ref{c}) ends up with a non-zero minimal momentum value. First, we expand Eq. (\ref{c})
with respect to $\beta $. We find

\begin{equation}
\left[ X,P\right] =i\hbar \left( 1+\frac{\beta }{2}X^{2}-\frac{\beta ^{2}}{8}%
X^{4}+\frac{\beta ^{3}}{16}X^{6}-...\right) .
\end{equation}%
Then, we obtain the uncertainty relation that arises in the MEUP as
\begin{eqnarray}
\left( \Delta X\right) \left( \Delta P\right) &\geqslant &\frac{\hbar }{2}%
\left \langle \sqrt{1+\alpha X^{2}}\right \rangle ,  \notag \\
&\geqslant &\frac{\hbar }{2}\left \langle 1+\frac{\beta }{2}X^{2}-\frac{%
\beta ^{2}}{8}X^{4}+\frac{\beta ^{3}}{16}X^{6}-...\right \rangle ,  \notag \\
&\geqslant &\frac{\hbar }{2}\left( 1+\frac{\beta }{2}\left \langle
X^{2}\right \rangle -\frac{\beta ^{2}}{8}\left \langle X^{4}\right \rangle +%
...\right).
\end{eqnarray}
We use the identity $\left \langle X^{2n}\right \rangle
\geqslant \left \langle X^{2}\right \rangle ^{n}$ \cite{HC1,HC2, SC, Po}, and we arrive at
\begin{eqnarray}
\left( \Delta X\right) \left( \Delta P\right) &\geqslant &\frac{\hbar }{2}%
\sqrt{1+\beta \left[ \left( \Delta X\right) ^{2}+\left \langle X\right
\rangle ^{2}\right] },  \label{C1}
\end{eqnarray}
up to the first order $\beta$. Therefore, we conclude that the absolutely smallest uncertainty in momentum is%
\begin{equation}
\left( \Delta P\right) _{MEUP}\simeq \hbar \sqrt{\frac{\beta }{2}}. \label{PMEUP}
\end{equation}

\section{Heuristic derivation of Unruh effect from modified uncertainty relations}

\begin{figure}[!htb]
	\begin{center}
\includegraphics[scale=1]{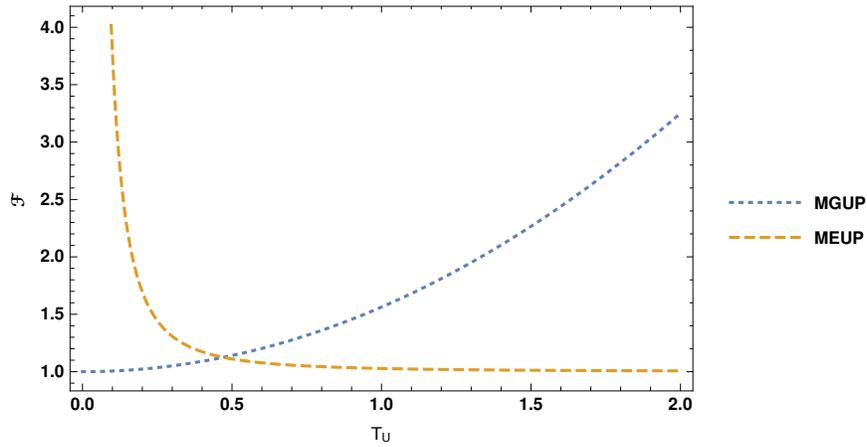}
\caption{ A comparison of the correction functions of both cases versus the Unruh temperature for the following numerical values $\hbar=c=K_B=1$ and $\alpha=\beta=0.5$. }
\label{fig1}\end{center}
\end{figure}

In this section, we extract the Unruh temperature starting from the
MGUP. We employ simple classical physics relations and combine them with the basic principles of
quantum mechanics, as done in  \cite{Scardigli,Jaume,Fabio}. This method allows us to estimate what kind of corrections are provoked by a modified uncertainty relations. We start by substituting $\Delta P=\frac{\Delta E}{c}$
in Eq.(\ref{b1}), we find%
\begin{equation}
\left( \Delta X\right) \left( \Delta E\right) =\frac{\hbar c}{2}\sqrt{1+%
\frac{\alpha }{c^{2}}\left( \Delta E\right) ^{2}}.
\end{equation}
We solve the quadratic equation for $\left( \Delta E\right)$, and  consider  the solution with positive sign. Then, we arrive at%
\begin{equation}
\left( \Delta E\right) =\frac{\hbar c}{2\left( \Delta X\right) \sqrt{1-\frac{%
\alpha \hbar ^{2}}{4\left( \Delta X\right) ^{2}}}}.  \label{d1}
\end{equation}
Then, by following \cite{Scardigli,Jaume,Fabio}, we adopt the distance along which each
particle must be accelerated in order to create $N$ pairs to  $\left( \Delta X\right) $%
\begin{equation}
\Delta X=\frac{2Nc^{2}}{a},  \label{d2}
\end{equation}
where $a$ is the acceleration of the frame. Next,  we interpret the energy fluctuation as a thermal agitation effect, thus, we
take%
\begin{equation}
\left( \Delta E\right) =\frac{3}{2}K_{B}T.  \label{d3}
\end{equation}
We use Eqs. (\ref{d2})
and (\ref{d3}) in Eq. (\ref{d1}). So that, we obtain an expression for the modified
Unruh temperature in the following form%
\begin{equation}
T_{MGUP}=\frac{T_{U}}{\sqrt{1-\frac{9}{4}\frac{\alpha }{c^{2}}%
K_{B}^{2}T_{U}^{2}}},
\end{equation}
where we fix $N=\pi /3$ and use $T_{U}=\frac{\hbar a}{2\pi K_{B}c}$.
Expanding $T_{MGUP}$ in power series over $\alpha $, we get%
\begin{equation}
T_{MGUP}\simeq T_{U}\mathcal{F}_{MGUP}\left( T_{U}\right) ,
\end{equation}%
while %
\begin{equation}
\mathcal{F}_{MGUP}\left( T_{U}\right) =1+\frac{9}{8}\frac{\alpha }{c^{2}}%
K_{B}^{2}T_{U}^{2}.
\end{equation}%
We observe that, in the MGUP approach, the correction function tends to increase the Unruh temperature. Next, we calculate the Unruh temperature in
the framework of MEUP which is encoded in the deformed algebra given in Eq. (\ref{c}). By following the same steps that are executed in the MGUP case, we obtain  the modified Unruh temperature  in the form of
\begin{equation}
T_{MEUP}=T_{U}\sqrt{1+\frac{c^{2}\hbar ^{2}\beta }{9K_{B}^{2}T_{U}^{2}}}. \label{CorMGUP}
\end{equation}%
Then, we expand it in power series of the deformation parameter and neglect the higher terms of
order $\mathcal{O}\left( \beta ^{2}\right) $. We obtain%
\begin{equation}
T_{MEUP}\simeq T_{U}\mathcal{F}_{MEUP}\left( T_{U}\right) ,
\end{equation}%
where
\begin{equation}
\mathcal{F}_{MEUP}\left( T_{U}\right) =\allowbreak 1+\frac{1}{18}\frac{\hbar
^{2}c^{2}\beta }{K_{B}^{2}T_{U}^{2}}. \label{CorMEUP}
\end{equation}%
Unlike the MGUP model, in the MEUP model the correction function tends to decrease the Unruh temperature. In order to demonstrate these behaviours of the correction functions  we depict Eqs. \eqref{CorMGUP} and \eqref{CorMEUP} versus the Unruh temperature in Fig. \ref{fig1}.
The MGUP and MEUP models are thought to be effective on the early and lately time of the universe dynamics, respectively \cite{Lutfu}. At a critical temperature value, $T_U^c=\frac{\hbar c}{3 K_B}\big(\frac{\beta}{\hbar^2 \alpha}\big)^{1/4}$, the correction functions have the same value,  $1+\frac{\hbar}{2}\sqrt{\alpha \beta}$.

\section{Black hole thermodynamics for the MEUP case}
\begin{figure}[!htb]
	\begin{center}
\includegraphics[scale=1]{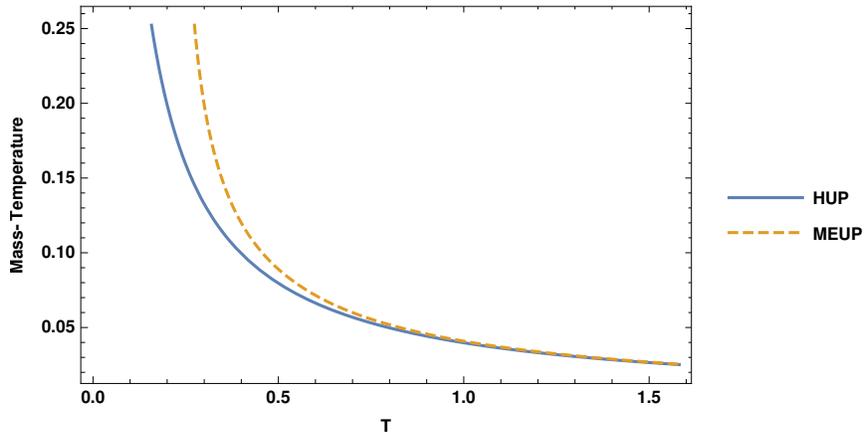}
\caption{A comparison of the mass-temperature functions for the following numerical values $G=\hbar=c=K_B=1$ and $\beta=0.2$. }
\label{fig2}
\end{center}
\end{figure}

We consider the line element of a four-dimensional Schwarzschild black hole that is given by%
\begin{equation}
ds^{2}=-\left( 1-\frac{2MG}{rc^{2}}\right) c^{2}dt^{2}+\left( 1-\frac{2MG}{%
rc^{2}}\right) ^{-1}dr^{2}+r^{2}d\Omega ^{2},
\end{equation}
where $M$ represents the mass of the black hole and $G$ is the Newton
universal gravitational constant. The Schwarzschild horizon radius, located
at $r_{s}$, is defined by%
\begin{equation}
r_{s}=\frac{2MG}{c^{2}},
\end{equation}%
For any massless quantum particle near the Schwarzschild black hole horizon
with mass $M$, its temperature can be written as%
\begin{equation}
T=\frac{c\left( \Delta P\right) }{K_{B}}. \label{E}
\end{equation}%
The minimal uncertainty in momentum, which is derived in Eq. \eqref{PMEUP}, leads to a lower bound temperature value for the black hole as
\begin{equation}
T_{0}=\frac{\hbar c}{K_{B}}\sqrt{\frac{\beta }{2}}\simeq 10^{-29}\mathrm{K}.
\end{equation}%
Next, we examine the effects of the MEUP scenario on the black hole mass, heat capacity and entropy functions. Hereafter, we follow the heuristic arguments of \cite{Scardigli,Jaume} and use the definition of the Schwarzschild black hole radius in the form of
\begin{equation}
\left( \Delta X\right) =2\pi r_{s}=\frac{4\pi MG}{c^{2}}. \label{E1}
\end{equation}
We employ Eqs.(\ref{E}) and (\ref{E1}) in Eq.(\ref{C1}), we find%
\begin{equation}
\left( \frac{4\pi MG}{c^{2}}\right) \left( K_{B}T\right) =\frac{\hbar c}{2}%
\sqrt{1+\beta \left( \frac{4\pi MG}{c^{2}}\right) ^{2}}.
\end{equation}%
This quadratic  mass-temperature equation has a solution of%
\begin{equation}
{M(T)}_{MEUP}= {M(T)}_{HUP}\left[{1-\frac{1}{2}\frac{T_{0}^{2}}{T^{2}}}\right]^{-1/2},
\label{E2}
\end{equation}%
where ${M(T)}_{HUP}=\frac{c^{2}}{8\pi G}\left( \frac{\hbar c}{K_{B}T}\right) .$
The mass-temperature function shows that the black hole mass is well-defined
for $T>\frac{T_{0}}{\sqrt{2}}$. Since $\beta$ is a small deformation parameter, thus $T_0$,  the black hole mass can be expanded as
\begin{equation}
{M(T)}_{MEUP}\simeq {M(T)}_{HUP}\left( 1+\frac{1}{4}\frac{T_{0}^{2}}{T^{2}}+\frac{3}{32}\frac{T_{0}^{4}}{T^{4}}+....\right) .
\end{equation}%
We note that when $\beta$ is taken as zero, hence $T_0=0$, the correction function value reduces to one, so that the mass-temperature function returns back to its ordinary form. To illustrate the modification function, we plot the mass-temperature functions of the two forms versus temperature in Fig. \ref{fig2}. We observe that the modification is effective only in a limited temperature interval. In higher temperatures, the MEUP effects can not be distinguished.

\begin{figure}[!htb]
	\begin{center}
\includegraphics[scale=1]{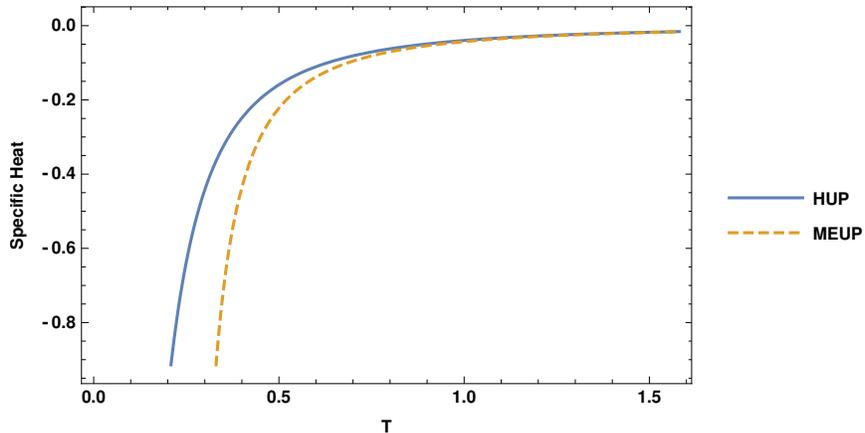}
\caption{A comparison of the specific heat functions for the following numerical values $G=\hbar=c=K_B=1$ and $\beta=0.2$. }
\label{fig3}
\end{center}
\end{figure}

Then, we obtain the temperature function in terms of mass as follows%
\begin{equation}
T_{MEUP}=\frac{\hbar c^{3}}{8\pi K_{B}MG}\sqrt{1+16\frac{\pi ^{2}\beta }{%
c^{2}}\frac{M^{2}}{M_{p}^{2}}}.
\end{equation}
Next, we investigate the specific heat function of the black hole. Here, we employ the following definition
\begin{equation}
C=c^{2}\frac{dM}{dT}.  \label{E3}
\end{equation}
By substituting Eq. (\ref{E2}) into Eq. (\ref{E3}), the specific heat
function of the black hole takes the form of
\begin{eqnarray}
{C(T)}_{MEUP}&=&{C(T)}_{HUP}\left[{1-\frac{1}{2}\frac{T_{0}^{2}}{T^{2}}}\right]^{-3/2},  \label{E4}
\end{eqnarray}
where ${C(T)}_{HUP}=-\frac{c^{4}}{8\pi G}\left( \frac{\hbar c}{K_{B}T^{2}}\right)
.$ The corrections to the standard Hawking specific heat are obtained by
expanding Eq. (\ref{E4}) in terms of $T_{0}$. So that, we find
\begin{equation}
{C(T)}_{MEUP}\simeq{C(T)}_{HUP}\left( 1+\frac{3}{4}\frac{T_{0}^{2}}{T^{2}}+\frac{15}{32}
\frac{T_{0}^{4}}{T^{4}}+...\right) ,  \label{E5}
\end{equation}%
It is worth noting that, for $\beta=0$ value we obtain
the same specific heat functions. We present their behavior versus the temperature in Fig. \ref{fig3}. Alike the mass-temperature functions, we observe the specific heat functions differ from each other only in relatively low-temperature values.

\begin{figure}[!htb]
	\begin{center}
\includegraphics[scale=1]{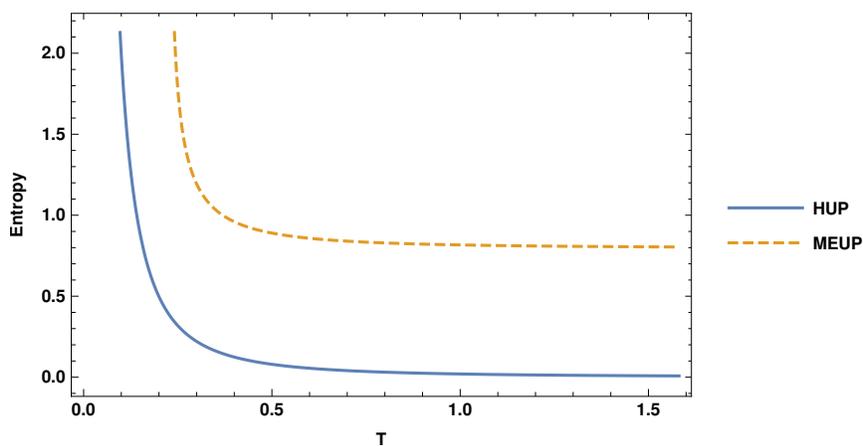}
\caption{A comparison of the entropy functions for the following numerical values $G=\hbar=c=K_B=1$ and $\beta=0.2$. For this MEUP scenario, the lowest temperature is $0.22$.}
\label{fig4}
\end{center}
\end{figure}

Next, we obtain explore the entropy function. We employ the black hole's mass temperature function, which is given in
Eq. (\ref{E2}), to derive the entropy in the usual fashion
\begin{equation}
S=\int \frac{dM}{T}=\int^{T}C\left( T\right) \frac{dT}{T}.  \label{E6}
\end{equation}%
We find%
\begin{equation}
{S(T)}_{MEUP}=\frac{c^{4}}{4\pi G}\left( \frac{\hbar c}{%
K_{B}T_{0}^{2}}\right) \left[{1-\frac{1}{2}\frac{T_{0}^{2}}{T^{2}}}\right]^{-1/2},
\end{equation}
while in the HUP case the entropy is
\begin{equation}
{S(T)}_{HUP}=\frac{c^{4}}{16\pi G}\left( \frac{\hbar c}{%
K_{B}T^{2}}\right). \
\end{equation}
We plot entropy functions in Fig. \ref{fig4}. We observe that a lower bound temperature value, that is $ \sqrt{2} T_0 / 2$, exists in the  MEUP approach, while not in the ordinary formalism not. Moreover, in the MEUP case entropy saturates in a non-zero value in high temperatures. In the absence of the correction term, that entropy value reduces to zero. Finally, we express them in terms of each other, and then we expand the correction function as
\begin{eqnarray}
{S(T)}_{MEUP}&=&{S(T)}_{HUP}\left( \frac{4T^2}{T_{0}^{2}}\right) \left[{1-\frac{1}{2}\frac{T_{0}^{2}}{T^{2}}}\right]^{-1/2}, \\
&\simeq&{S(T)}_{HUP}\bigg[ \frac{4T^2}{T_{0}^{2}}+1+\frac{3}{8}\frac{T_{0}^{2}}{T^2}+\cdots\bigg]. \,\,\,\,\,\,
\end{eqnarray}

\section{Black hole thermodynamics for the MGUP case}

\begin{figure}[!htb]
	\begin{center}
\includegraphics[scale=1]{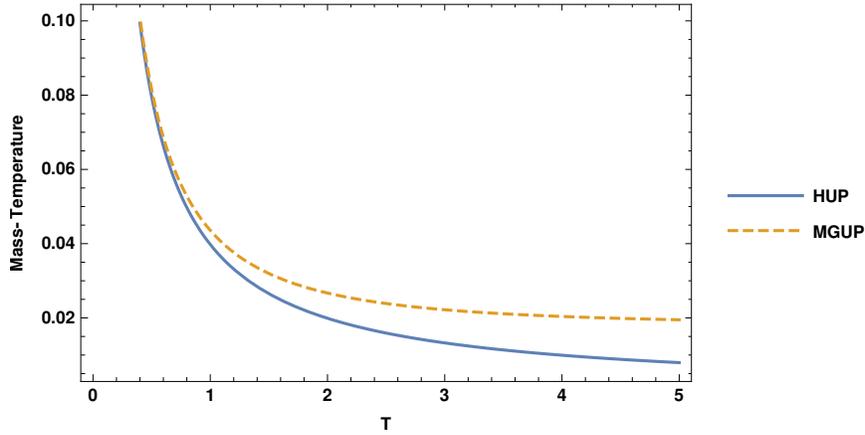}
\caption{A comparison of the mass-temperature functions for the following numerical values $G=\hbar=c=K_B=1$ and $\alpha=0.2$. }
\label{fig5}
\end{center}
\end{figure}

Near-horizon geometry considerations suggests to set $\Delta X=2\pi r_{s}$ \cite{Adler}.
Therefore,  Eq. (\ref{b1a}) leads to  a minimum horizon radius and a minimum mass values for a black hole in the form of%
\begin{equation}
r_{s}\simeq \frac{\hbar }{2\pi }\sqrt{\frac{\alpha }{2}},\text{ \  \  \ }%
M_{0}=M_{p}^{2}\frac{c}{4\pi }\sqrt{\frac{\alpha }{2}},
\end{equation}%
which mean black holes with mass smaller than $M_0$ does not exist. Using Eqs. (\ref{b1a}), (\ref{E}), and (\ref{E1})  we determine the MGUP black hole mass as%
\begin{equation}
{M(T)}_{MGUP}={M(T)}_{HUP}\sqrt{1+\alpha\frac{K_{B}^{2}T^{2} }{c^{2}}}.  \label{U}
\end{equation}%
The MGUP black hole does not propose a lower bound on the
temperature unlike the MEUP black hole. We plot the mass-temperature functions in Fig. \ref{fig5} to present the characteristic behaviors. We observe that the black hole mass decreases with temperature, however in the MGUP case this decrease is in a slower rate. Moreover, in the MGUP case, at high temperatures the mass-temperature function saturates at  a non-zero value, unlike the HUP case.

\begin{figure}[!htb]
	\begin{center}
\includegraphics[scale=1]{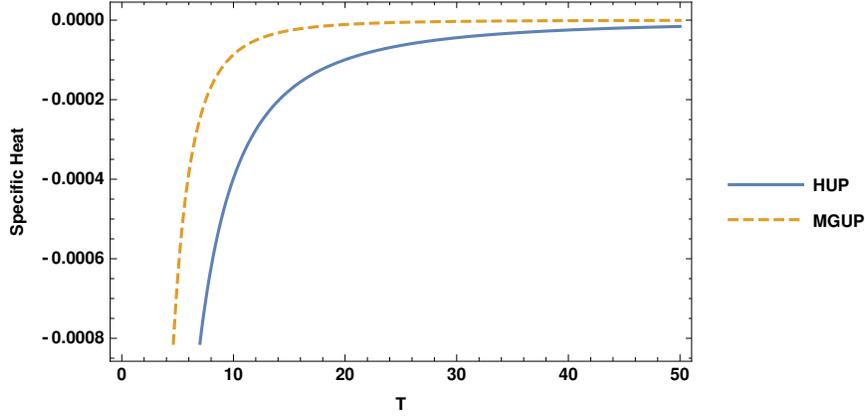}
\caption{A comparison of the specific heat functions for the following numerical values $G=\hbar=c=K_B=1$ and $\alpha=0.2$. }
\label{fig6}
\end{center}
\end{figure}

\begin{figure}[!htb]
	\begin{center}
\includegraphics[scale=1]{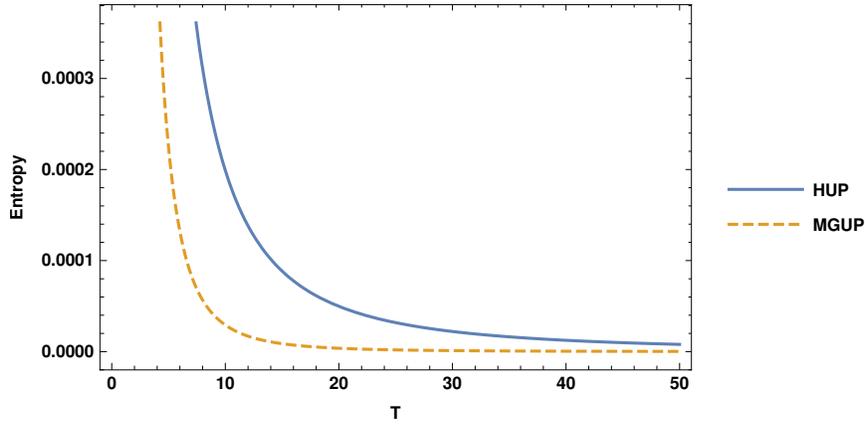}
\caption{A comparison of the entropy functions for the following numerical values $G=\hbar=c=K_B=1$ and $\alpha=0.2$. }
\label{fig7}\end{center}
\end{figure}

Then, we expand the correction function of  Eq. (\ref{U}) in terms of $\alpha $ deformation parameter. We find
\begin{eqnarray}
{M(T)}_{MGUP}\simeq{M(T)}_{HUP}\Big( \allowbreak 1+\frac{\alpha}{2}\frac{ K_{B}^{2}T^{2}}{c^{2}%
}++...\Big) ,
\end{eqnarray}%
Here, the first term is ordinary Hawking mass term while the second term is the first order correction
brought by MGUP. It is interesting to inverse Eq. (\ref{U}) and
express the temperature of the black hole as a function of the mass%
\begin{equation}
T(M)=\frac{\frac{M_{p}^{2}c^{2}}{8\pi K_{B}M}}{\sqrt{1-\frac{M_{0}^{2}}{2M^{2}}}%
}. \label{TMGUP}
\end{equation}%
Eq. \eqref{TMGUP} produces a complex number value  when $M$ falls below $\sqrt{2}M_{0}/2.$
We see that the evaporation ceases at about the minimum mass $M_{0}$,
leading to stable relics. So "the MGUP" stabilizes the ground state
of a black hole just as "the HUP" stabilizes the ground state of a hydrogen atom. Next, we determine the MGUP black hole specific heat.
Inserting Eq.(\ref{U}) into  Eq.(\ref{E3}), we get
\begin{equation}
{C(T)}_{MGUP}={C(T)}_{HUP}\bigg[1+\alpha\frac{K_{B}^{2}T^{2} }{c^{2}}\bigg]^{-1/2}.
\end{equation}%
In Fig. \ref{fig6}, we compare the specific heat functions. We observe that in the MGUP case, it reaches saturation value faster than the HUP case. Indeed, the specific heat function behaves like inverse cube and inverse square  in MGUP and HUP cases in higher temperatures, respectively.
Since the deformation parameter is very small, we expand the specific heat function up to the first order of $\alpha$. We find
\begin{equation}
{C(T)}_{MGUP}\simeq{C(T)}_{HUP}\left( 1-\frac{\alpha }{2}\frac{K_{B}^{2}T^{2}}{c^{2}}+...\right) .
\end{equation}%
Finally, we examine the MGUP black hole entropy function. We follow the same steps of the previous section and obtain the entropy function as follows
\begin{eqnarray}
\frac{{S(T)}_{MGUP}}{K_{B}}&=&4\pi \left( \frac{M_{HUP}}{M_{p}}\right) ^{2}\sqrt{1+\frac{
\alpha }{c^{2}}K_{B}^{2}T^{2}}\nonumber \\
&\times&\left[ 1-\frac{\alpha K_{B}^{2}T^{2}}{c^{2}}
\frac{\arctan\frac{1}{\sqrt{1+\frac{\alpha }{c^{2}}K_{B}^{2}T^{2}}}}{
\sqrt{1+\frac{\alpha }{c^{2}}K_{B}^{2}T^{2}}}\right].\,\,\,\,\,\,\,\,\,\,\,\,
\end{eqnarray}
We depict the entropy functions in Fig. \ref{fig7}. We observe that the entropy of the MGUP black hole tends to zero faster than the HUP case black hole.

\section{Conclusion}
There are several alternatives to construct a  noncommutative model in momentum or position space by considering different isotropic parameterizations. In this paper, we proposed two new forms of noncommutative models, namely MGUP and MEUP,  and opened a new window to the field by investigating the modifications caused by these models on the Unruh temperature and the thermodynamics of the Schwarzschild black hole.

Regarding the Unruh temperature,  we found that in the MEUP model the correction function tends to decrease the Unruh temperature, while in the MGUP model the correction function tends to increase the Unruh temperature. Within these two scenarios, we observed that the thermodynamic functions of a Schwarzschild black hole, i.e. specific heat and entropy, affect differently. In particular,
the expansion of the entropy function in terms of the deformation parameters do not contain a logarithmic term which
appears in the GUP-corrected Schwarzschild black hole case.
Similarly, we found that the mass temperature functions also vary in different characters. Moreover, we showed that the minimal uncertainty in momentum leads to a lower bound for the black hole temperature while the minimal length leads to a black hole minimum horizon radius and minimum mass.

\end{document}